\documentstyle[aaspp4,psfig]{article}
\def\ap{$\sim$}
\def\ges{$\;_\sim^>\;$}
\def\les{$\;_\sim^<\;$}
\def\deg{$^\circ$}
\def\etal{{\it et al.}}
\def\g{$\gamma$-ray}
\def\min{$^{\prime}$}
\def\pt#1{$10^{#1}$}
\def\tpt#1{$\times10^{#1}$}
\def\cms{cm$^{-2}$s$^{-1}$}

\begin{document}
To be published in Ap J,  May 20, 1997 issue, Vol. 481.

\title{The Identification of EGRET Sources with Flat-Spectrum Radio Sources}
\author{J. R. Mattox\altaffilmark{1,2},
J. Schachter\altaffilmark{3}
L. Molnar\altaffilmark{4},
R. C. Hartman\altaffilmark{5},
A. R. Patnaik\altaffilmark{6}
}
\altaffiltext{1}{Astronomy Department, Boston University, 725 Commonwealth Ave.,
Boston, MA 02215} 
\altaffiltext{2}{Former affiliations: Astronomy Department, University Of Maryland;
Universities Space Research Association;
Compton Observatory Science Support Center, NASA/GSFC;
Gamma-Ray Astrophysics Branch, NASA/GSFC}

\altaffiltext{3}{Center for Astrophysics, Cambridge, MA 02138}
\altaffiltext{4}{Dept. Of Phys. \& Astron, University of Iowa, Iowa City, IA 52242}
\altaffiltext{5}{Code 662, NASA/GSFC, Greenbelt MD 20771}
\altaffiltext{6}{Max-Planck Institut f\"ur Radioastronomie, Bonn, GERMANY }

\begin{abstract}
We present a method to assess the reliability
of the identification of EGRET sources with extragalactic radio sources.
We verify that EGRET is detecting the blazar class of AGN. However,
many published 
identifications are found to be questionable.
We provide a table of 42 blazars
 which  we expect to be robust identifications of EGRET sources. This includes
one previously unidentified EGRET source,
the lensed AGN PKS 1830$-$210 near the direction of the Galactic center.
We provide the best available positions
for 16 more radio sources which are also potential identifications for
previously unidentified EGRET sources.
All high Galactic latitude EGRET sources ($|b|>3^\circ$) 
which demonstrate significant 
variability can be identified with flat spectrum radio sources. 
This suggests that
EGRET is not detecting any  type of AGN other than blazars. This
identification method has been used to establish with 99.998\% confidence
that the peak \g\ flux of a blazar is correlated with its
average 5 GHz radio flux. An even better correlation is seen between 
\g\ flux and the 2.29 GHz
flux density measured with VLBI at the base of the radio jet.
Also, using high confidence identifications,
we find that the radio sources identified with EGRET sources
have larger correlated VLBI flux densities than the parent population
of flat radio spectrum sources.
\end{abstract}

\keywords{Gamma Rays: observations
--- galaxies: active --- quasars: general}

\vskip 1cm
\centerline{\bf 1. Introduction}
The Energetic Gamma Ray Experiment Telescope (EGRET) aboard the Compton
Gamma Ray Observatory is sensitive in the energy range 30 MeV to 30
GeV (Thompson \etal\ 1993). 
A catalog of EGRET sources based on the
first 30 months of exposure is given by Thompson \etal\ (1995, henceforth
TH95).
It includes 129 sources.
Five are pulsars, 40 are listed as ``high-confidence identifications''
of AGN, and 11 are listed as ``lower-confidence identifications''
of AGN. 
A total of 71  are unidentified.

We demonstrate quantitatively in this paper that the claim 
that EGRET is detecting the \g\ emission of some members of
the blazar class of AGN is valid
(see von Montigny \etal\ 1995 for a compendium
of EGRET blazar results for the first \ap1/2 of the EGRET mission).
By blazars, we mean AGN with
strong, compact, flat-spectrum ($\alpha$ be \ges $-0.5$, 
where $S(\nu)\propto\nu^{\alpha}$) 
radio emission; and which also
show continuum domination of the optical emission, and/or
significant optical polarization, and/or 
significant changes in optical flux on short time scales
(designated Optically Violently Variable, or OVV).
These observational properties are believed to result from emission by material
in a relativistic  jet which is directed within \ap10\deg\ of the
line of sight.
The blazar class includes objects classified as BL Lacertae type objects,
high polarization quasars (HPQ), and OVV quasars.
The  apparent \g\ luminosity is as much as one hundred times larger than that at
all other wavelengths for some flaring EGRET blazars. 
Variability of the \g\ flux
from some blazars on a time-scale as short as 4 hours 
(Mattox \etal\ 1997) implies that the
\g s are emitted from a compact region. 
The \g\ emission is not understood.
It may offer fresh insight into the blazar phenomenon, and useful diagnostics. 

A careful analysis of the EGRET source identifications is in order because
the source of the \g\ emission is not well located.
EGRET position estimates are  imprecise 
because of the wide point spread function of the EGRET instrument.
The half-angle of a cone which contains
68\% of the EGRET events from a point source at a  specific energy is
well fit (Mattox \etal\ 1996a) by
\begin{equation}
\theta_{68}=5.85^\circ \lbrack{E_\gamma /100 \rm{MeV}}\rbrack^{-0.534}
\end{equation}
The localization expected for a source is approximately
$\theta= \theta_{PSF}/\sqrt N$
where $N$ is the number of \g s detected from the source
(Thompson 1986). For weak 
EGRET sources, the
95\% confidence position error ellipses are as large as \ap5
square degrees. 

In the first EGRET catalog (Fichtel \etal\ 1994), EGRET blazars were classified
as ``positive detections'' or ``marginal detections'' depending on whether
the significance of point-source \g\ emission exceeded 5$\sigma$ or
4$\sigma$ respectively. The confidence of the identification was not explicitly
addressed. The second EGRET catalog (TH95) did explicitly address
the confidence of  identification.
TH95 classified an identification as ``high-confidence''
if the radio source was located within the 95\% confidence contour of the
EGRET position estimate. Also, TH95 classified
some radio sources beyond the 95\% confidence contour as ``lower-confidence''
identifications. The 5 GHz flux density ($S_5$) of a radio counterpart
was generally required by TH95 to be \ges 1 Jy; and the 
radio spectral index to be generally $\alpha$\ges$-0.5$. 
No formal consideration was given  to the size of the EGRET
error region and where the radio source lay within that error region
(other than whether  the source was located within 
 the 95\%  confidence contour of the EGRET position estimate
region or not). No formal consideration was given  to 
the number density of potentially confusing
radio sources (other than a loose requirement that $S_5$ be \ges 1 Jy).  
No formal consideration was given  to the  radio spectral index
(other than a loose requirement that $\alpha$ be \ges $-0.5$). 
No consideration was given to the {\it a priori} probability of
detecting a radio source by EGRET ---
it is  less than unity because
many blazars which are bright at
other wavelengths have been observed by EGRET without being detected.
Also, the decision by TH95 on which radio sources beyond the 95\% confidence contour
to include as ``lower-confidence''
identifications was subjective.

We have developed an analysis of EGRET radio source identification which
quantitatively incorporates all of this information.
We avoid a cutoff in the search for an identification
at an arbitrary confidence contour of the EGRET position estimate
region.
Also, we  avoid  terminating the search for candidates
at a specific minimum radio flux density or radio spectral index. 
Bayes' theorem is used to incorporate 
the {\it a priori} probability of
detecting a radio source 
(see Sturrock 1973 and Loredo 1990 for discussions of the use of
Bayesian statistics in astronomy). It is reasonable to use a Bayesian analysis
here for the following reasons: 1) we can demonstrate  
with very high confidence  (see \S 3.1)
that EGRET is detecting flat-spectrum radio sources
without making any {\it a priori} assumptions;
2) we can derive a reasonable estimate for the {\it a priori} 
probability of EGRET detecting a radio source as a function of radio flux
(see \S 3.2).

The method is described
in \S 2. In \S 3 we  demonstrate conclusively that
EGRET is detecting blazars and obtain an estimate for the {\it a priori} 
probability of EGRET detecting a radio source as a function of radio flux; and
then provide a quantitative study of the reliability of the 
EGRET identifications of TH95 and Thompson \etal\ (1996), and  
search for additional radio
identifications among the previously unidentified EGRET sources.
Limitations of this analysis are discussed in \S 3.6.
Implications are discussed in
\S 4. In \S 3.3.1, we use the high probability identifications
 of \S 3.3 to demonstrate
that the EGRET blazars have higher  VLBI flux densities than 
the parent population of flat-spectrum radio sources. 

\vskip 1cm
\centerline{\bf 2.  EGRET Source Identification}

We analyze the probability of a correct identification of an
EGRET source with a radio source. The calculation considers
the  size of the EGRET
error region and where the radio source lies within that error region.
Bayes' theorem is used to include consideration of 
the number density of potentially confusing
radio sources and the 
{\it a priori} probability of
detecting a specific radio source by EGRET.
This method is an extension of that
used by de Ruiter, Willis and Arp (1977)
to identify optical counterparts of radio sources.
This method has also been applied to identification of IRAS sources
(Wolstencroft \etal\ 1992) and X-ray sources
(Schachter, \etal\ 1997).

\vskip 1cm
\centerline{\bf 2.1 Radio Catalogs}

Because blazars are 
flat-spectrum radio
sources, high frequency surveys are appropriate for this study
since they efficiently detect
flat spectrum sources.
We have used the 4.85 GHz Greenbank (GB) survey (Condon,
Broderick, \& Seielstad 1989) for the northern sky and  the 4.85 GHz
Parkes--MIT--NRAO (PMN)
survey (Griffith \& Wright 1993) for the southern sky.
Both surveys have a threshold flux density of \ap30 mJy and
give positions with \ap20\min\min\ uncertainty.
They
are both  confusion limited at Galactic latitudes less than 3\deg.
Therefore, we do not attempt to identify the 19 unidentified
EGRET sources at latitudes
less than 3\deg\ in this paper. Also, the PMN survey is confusion limited 
near the Large Magellanic Cloud and Centaurus A. For this reason,
EGRET sources 2EG J0532-6914 and 2EG J1324-4317
are also excluded from this analysis.

The GB survey covered the declination range 0\deg$<\delta<$75\deg.
We have used the catalog of Becker, White, and Edwards (1991) of
53522 sources derived from this survey.
The White and Becker (1992) catalog
of sources from the 1.4 GHz GB survey was used to
obtain an $\alpha_{1.4-4.85}$ spectral index for the 22384
sources which are found in both catalogs.
Because the 1.4 GHz survey preceded the
4.85 GHz survey by \ap3 years, variability
of blazar type AGN can result in this index differing by as much as
\ap1 from what would be obtained with simultaneous
measurements. 

The PMN survey (Griffith \& Wright 1993)
covered the declination range $-87.5$\deg$<\delta<$10\deg.
There are small gaps in the survey due to observational problems.
The PMN catalog 
contains 49329 sources with $S_5> 30$ mJy.
Spectral indices are provided for the \ap10\% of the PMN sources
which were detected in the Parkes 2.7 GHz survey
(Bolton, Wright, and Savage 1979). 
Again, variability can cause a substantial
error in this spectral index.
Only one unidentified source (2EG J1332+8821) falls completely outside of these
surveys. The error ellipses of two EGRET sources  (J1409$-$0742 and J1513$-$0857) 
fall partially in gaps in the PMN survey, but both have strong identifications
from the  PKS catalog.
We retain the original eight digit position names of both catalogs.
The GB source names are prefixed with ``B''
because the position names are based on B1950 celestial coordinates.
The PMN source names are prefixed with ``J''  corresponding to
J2000 celestial coordinates. We use the GB survey in the region
where the surveys overlap, 0\deg$<\delta<$10\deg.

\vskip 1cm
\centerline{\bf 2.2 Probability of Correct Identification}

For a radio source found at an angle  $r$ from the center of
an EGRET error ellipse, there are two possibilities:
the source is the radio counterpart of the EGRET source;
or, the source is a confusing source which is close to the EGRET
source by accident. We denote the probabilities
of these two hypotheses by  $p(id|r)$ and $p(c|r)$ respectively.

If the radio source is merely a 
confusing source, the Poisson distribution pertains.
The probability that one or more radio sources are accidently within  the
angle $r$ is
\begin{equation}
p(r|c)=1 - e^{- r^2/r_0^2},
\end{equation}
where $r_0$ is the characteristic angle between confusing sources,
\begin{equation}
r_0 = (\pi \rho(S_5,\alpha))^{-{1\over2}},
\label{ro}\end{equation}
and $\rho(S_5,\alpha)$ is
the number density of radio sources
with at least the flux density of the candidate radio source, and a spectrum
at least as flat.
The differential probability is
\begin{equation}
dp(r|c)=2{r\over r_0^2} e^{- r^2/r_0^2}dr
\label{conf-dist}
\end{equation}
For sources without spectral indices, we
assume a spectral index of $-0.5$. 
This results in a
conservative estimate of the significance of the identification
for radio  sources which are indeed flat spectrum. A refined estimate of
the probability of correct identification can be made after further radio
observations yield a spectral index.
We obtain $\rho(S_5,\alpha)$ from the
GB survey for the 16300 square degree region 0\deg$<\delta<75$\deg,
and $|b|>10$\deg.
The integral of $\rho(S_5,\alpha)$
over $\alpha$ is observed to have a Euclidean slope between 100 and
1000 mJy in agreement with the result of Kellerman and Wall (1987).
We use this  number density also for analysis of PMN counterparts.

If the radio source is the counterpart of an EGRET source,
it is appropriate to use the EGRET data
to estimate the probability distribution, $p(r|id)$, of the
angle $r$. 
A likelihood analysis of the EGRET data (Mattox \etal\ 1996a)
yields the 95\% confidence position contour from the likelihood
ratio test using Wilks' theorem.
These contours are usually fit accurately with ellipses. The
ellipse is centered
on the centroid of the 95\% confidence location region which is given
by TH95 as the position estimate. This position is observed to differ by only
a small amount from the position of maximum likely which is formally the
correct position estimate.
The ellipse is
specified by the  semi-major axis $a$, semi-minor axis $b$,
and the  position angle of the major axis $\phi$.
The departure from circularity is often substantial. This results from
source confusion (nearby \g\ sources below the EGRET detection threshold), 
error in the model of Galactic diffuse \g\ emission, 
and statistical fluctuation.
The elliptical fits  given by TH95
were used in our analysis. 
The parameters for the elliptical fits which are not given in
TH95 (because of an imprecise fit) have also been used
in this analysis:
2EG J0426+6618,     $a=$66\min,    $b=$54\min,    $\phi=$27\deg;
2EG J0511+5523,     $a=$135\min,    $b=$36\min,    $\phi=$24\deg;
2EG J0545+3943,     $a=$48\min,    $b=$47\min,    $\phi=$94\deg;
These fits are
all accurate to within \ap30\% in radius which is  not substantially worse
than the potential effect of error in the Galactic diffuse model and 
undetected nearby \g\  sources.

For strong sources, we observe that the dependence  of the
log of the likelihood of the EGRET
data on the assumed source position is well represented by a paraboloid. 
The 95\%
confidence contour corresponds to a drop in the log of the
likelihood of 3.00 from the
maximum (Mattox \etal\ 1996a). Therefore, we represent
the log likelihood surface corresponding to an elliptical error region
as an elliptic paraboloid
$\ln L_{max}-\ln L=3r^2/\Psi^2$, where
\begin{equation}
\Psi=\lbrack a^{-2}\cos^2(\theta-\phi)+ b^{-2}\sin^2(\theta-\phi)\rbrack^{-1/2},
\label{cont}
\end{equation}
and $\theta$\ is the position angle.
For a circular error region, $\Psi$ is simply the 95\% confidence radius.
${}$From Wilks' theorem, $2(lnL_{max}-lnL)$ is  distributed as $\chi^2_2$
in the null hypothesis (Mattox \etal\ 1996a).
Thus, the
confidence with which a source may be expected within a region
delineated by
a $\Delta\equiv \ln L_{max}- \ln L$ decrease in log likelihood is
\begin{equation}
C=1-\int_{2\Delta}^\infty\chi^2_2(\zeta)d\zeta=1-e^{-\Delta}
\label{con}
\end{equation}
Therefore, the probability of the EGRET source being located at an angle smaller
than $r$ is
\begin{equation}
p(r|id)=1-e^{-3r^2/\Psi^2}
\end{equation}
and
\begin{equation}
dp(r|id)=6{r\over\Psi^2}e^{-3r^2/\Psi^2}dr
\label{id-dist}
\end{equation}

Bayes' theorem allows the observed 
angle between the radio source and the EGRET position estimate, $r$, 
to be used with an {\it a priori} 
probability and
the expected distributions of $r$ under the two possible hypotheses (equations
\ref{conf-dist} and \ref{id-dist}) to obtain $p(id|r)$,
the {\it a posteriori}
probability that a radio source is the correct identification of an EGRET
source.
\begin{equation}
p(id|r)={p(id)dp(r|id)\over dp(r)}=
{p(id)dp(r|id)\over p(id)dp(r|id) + p(c)dp(r|c)}
\end{equation}
With the substitution $\eta\equiv p(id)$ for the {\it a priori} probability,
and using the fact that one of the two hypotheses is true, 
either a source is the identification or
it is a confusing source, $p(c)=1-p(id)$;
we obtain
\begin{equation}
p(id|r)={{\eta\over1-\eta}LR\over {\eta\over1-\eta}LR+1}
\label{bayes},\end{equation}
where the likelihood ratio, $LR$, is defined to be
\begin{equation}
LR\equiv {dp(r|id)\over dp(r|c)}=3{r_0^2 \over \Psi^2}
e^{-r^2(3/\Psi^2-1/r_0^2)}
\label{lr}\end{equation}
Figure 1 shows the $LR$ and $\eta$ dependence of $p(id|r)$ given by
equation \ref{bayes}.

\begin{figure}
\hbox{\psfig{figure=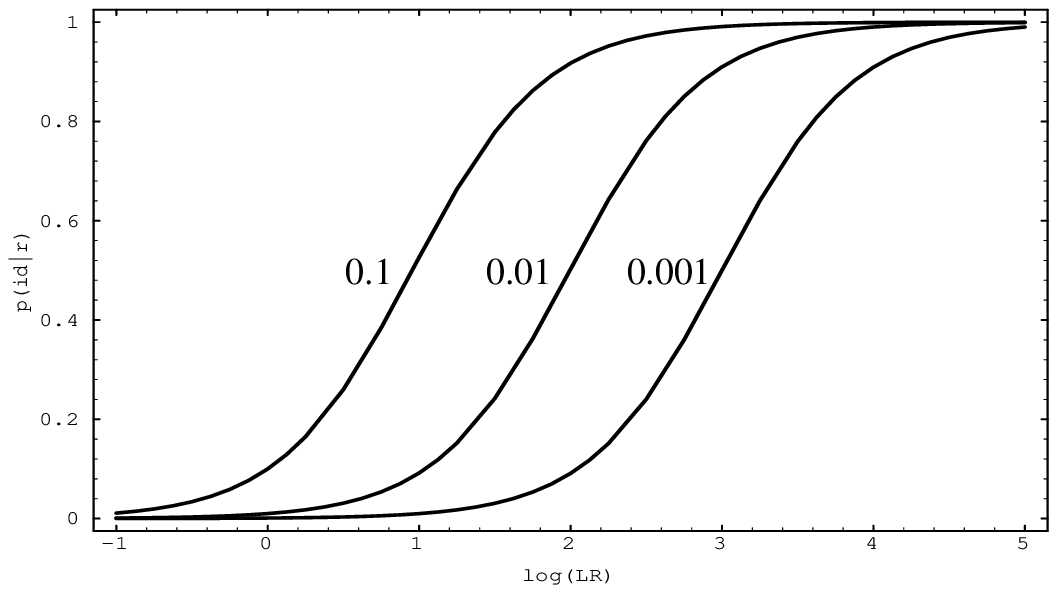,width=3.in,angle=270}}
\caption{
\baselineskip=11pt 
The relationship of equation (3) for the
probability of a correct identification
$p(id|r)$ as a function of the log of $LR$
 for the indicated values
of $\eta$.
} 
\end{figure}

We note that $LR$ diverges at large $r$ if
$\Psi>\sqrt3\ r_0$. 
In this case, the differential probability of $r$ 
under the confusion hypothesis falls off more rapidly with $r$ than it
does under the identification hypothesis.
This reflects the fact that the
size of the EGRET error region, $\Psi$, is larger than the 
characteristic angle between confusing sources, $r_0$. A meaningful
identification is not possible. We note that
$LR<1$ at $r=0$ in this case.
Therefore, we require
$\Psi<\sqrt3\ r_0$ before we consider a radio source as a candidate
identification of an EGRET source.
This limits the number
density of sources for which an attempted identification is meaningful. The
consequence is that a 5 GHz flux density of \ges50 mJy is required
for a source to be a meaningful candidate for identification given the
error ellipse sizes typical of the unidentified EGRET sources. 

\vskip 1cm
\centerline{\bf 3. Results}

Table 1 presents the results of the analysis described in 
\S2 for the 102 EGRET sources from TH95 which are not
identified as pulsars, and which are not in regions where the radio
surveys are source confused ($|b|< 3$\deg, and near the LMC and Cen A).
All radio sources are given  for which an {\it a posteriori}
probability, $p(id|r)>0.001$ is found.
The radio source with the
highest value of $p(id|r)$ is included for all EGRET sources 
which fall within the radio survey regions.
We note that sources with radio flux densities as small as \ap50 mJy
appear, but with very small
probabilities of being EGRET sources. The same analysis is presented in 
Table 2 for the sources in the Supplement to the Second EGRET Catalog 
(Thompson \etal\ 1996).
The content of Tables 1\& 2 is:
\begin{itemize}
\item[]
Column (1)  contains the J2000 position name of the EGRET
source from TH95 or Thompson \etal\ (1996).

\item[]
Column (2) is the Galactic position of the EGRET source (from TH95
or Thompson \etal\ 1996).
For the five sources where TH95 accidently gave the radio position
(0528+134, Mkn 421, 3C 279, 1406-076, 3C 454.3),
the EGRET positions have been substituted.

\item[]
Column (3) is the variability index  
which is $-{\rm log}_{10}(P)$, where
$P$ is the probability that the source is invariant based on a
$\chi^2$ test of flux for all available viewing periods
(McLaughlin \etal\ 1996). A larger variability index
corresponds to a higher probability that the source is variable.

\item[]
Column (4) is the identification   given (for some sources)
by TH95. A question mark following the B1950
radio position name means that TH95 
indicated that it was a 
``lower-confidence identification''; otherwise, the identification was
indicated by TH95 to be a ``high-confidence identification''.

\item[]
Column (5) is the name of the potential radio counterpart.  
The original  B1950 position names are used for Greenbank sources, and
the original J2000 position names are used for PMN sources.

\item[]
Column (6) is a more common name for some potential
radio counterpart.

\item[]
Column (7) is the 4.85 GHz   flux density in mJy from the survey catalog.  

\item[]
Column (8) is the radio spectral index, $\alpha$, from the radio
survey flux densities if available,
$S(\nu)\propto\nu^{\alpha}$.

\item[]
Column (9) is $r_0$,  the characteristic angle (in arcmin)
between confusing sources
which are at least as bright and flat as this radio source
(see equation \ref{ro}).

\item[]
Column (10) is $\eta=p(id)$,  the {\it a priori} probability  that the radio source
is a \g\ source from equation \ref{eta}.

\item[]
Column (11) is the angle, $r$ (in arcmin), between the 
 EGRET position estimate and the radio position.

\item[]
Column (12) is $\Psi$, the radius 
(in arcmin) of the 95\% confidence contour in the direction
of the radio source from equation \ref{cont}.

\item[]
Column (13) is the 
position confidence contour at the radio position (from equation \ref{con}).    

\item[]
Column (14) is the likelihood ratio (equation \ref{lr}) indicating the
strength of the indication for the identification.

\item[]
Column (15) is the {\it a posteriori}
probability that the identification is correct
as determined with equation \ref{bayes}.
\end{itemize}

\vskip 0.5cm
\centerline{\bf 3.1 The Statistical Significance of the EGRET
Detection of Radio Loud AGN}

We can now address an important question: with what confidence can we
conclude that EGRET is detecting radio loud AGN?
At this point, it is not appropriate to use
Bayesian statistics.
We have not yet  determined 
the {\it a priori} probability
of a flat spectrum radio
source being an EGRET source.
As illustrated by Figure 1,  
the {\it a posteriori} probabilities
depends very strongly on the {\it a priori} probability.
Note that for 
{\it a priori} probabilities
$\eta=1$ or $\eta=0$, equation \ref{bayes} shows that
Bayes' theorem implies that
the EGRET data do not influence the expectation about whether or not a
radio source emits \g s.

Since Bayesian statistics 
are of no use at this point, 
we apply a ``frequentist'' ansatz, a Monte Carlo calculation.
We use $LR$ (equation \ref{lr}) as an empirical indication of the strength of
a potential identification. We find that the distribution of $LR$ 
for the most likely identification of each source in Table
1 differs dramatically from the distribution found by Monte Carlo calculation
assuming no EGRET  detections of radio loud AGN.
Simulated EGRET source positions were chosen randomly with $|b|>3$\deg\ and 
used with the error ellipse parameters of TH95.
The analysis of \S2 for \ap3000 simulated sources
yielded an estimate for
the distribution of $LR$ in the null hypothesis.
A long tail at large $LR$ was observed. 
Confusing radio sources with $LR>$\pt{3}
were found at a frequency of $0.010\pm0.002$.
This corresponds to the occurrence at a non-negligible frequency
of spurious coincidences which
look fairly solid.
We find no strong dependence of  $LR$  on the size of the
error ellipse, supporting the validity of the method we describe in \S 2.

The strongest identification (and coincidently the first, Hartman \etal\ 1992)
is 3C 279 with $LR=4.6$\tpt5.
The large value of $LR$ reflects the fact that 3C 279 is centrally located in
a small  error ellipse, and  that
sources as bright and flat as 3C 279 are very rare:
$\rho(S_5\ge17 {\rm Jy},\alpha\ge0.3)=
$\ 4\tpt{-3} per square degree.
The largest observed $LR$ in the \ap3000 Monte Carlo
trials 
was 6\tpt3. Therefore, the 3C 279 detection alone
indicates that AGN are  seen by EGRET with a confidence
well above (1-102/3000)=97\%.
With the admission of an {\it a priori}
probability as small as $\eta=$\pt{-3}, equation \ref{bayes} 
implies an {\it a posteriori}
probability of 0.998 that 3C 279 is the EGRET source counterpart; and
with the self-consistent {\it a priori}
probability of $\eta=0.2$ which we derive in \S3.2, the 
{\it a posteriori}
probability that the identification is correct is 0.999991.

Even stronger evidence that
EGRET is detecting radio loud AGN
is obtained by  considering  all
sources in Table 1.
${}$From the Monte Carlo calculation, $LR>100$ occurs with a frequency of $0.05\pm0.004$
in the null hypothesis.
In comparison, 32 of the 102 EGRET sources in Table 1 have
an identification with $LR>100$.
The probability to obtain this by chance is
\begin{equation}
\sum_{n=32}^{102} 0.95^{(102-n)} 0.05^{n} {102!\over(102-n)! n!}=\ 2\times10^{-17}
\label{fact}\end{equation}
This is a very conservative estimate of the significance because
many of the values of $LR$ are much larger than 100.
However, there are ``trials'' made in the selection of the type of source
to try to identify with EGRET sources. The number of such trials cannot
exceed the \ap100 independent types of astronomical objects.
There are also possible  ``trials'' made in the selection of
the energy range (e.g. 2022$-$077) or diffuse  model (e.g. 1622$-$253)
which was used to obtain an EGRET position for an identification.
This number is \les 2.
The choice of the minimum value of $LR$ of 100 for this analysis
 also corresponds to 
a number of ``trials'', but the number
 is less than \les 10.
Also,
the use of the radio spectral index in the determination of the
number density of radio sources can be regarded as a trials factor
of size \ap2. (We demonstrate in \S 4.2 that EGRET is generally not
detecting  steep-spectrum radio sources, 
so the use of the radio spectral index 
in the determination of the
number density of radio sources is justified.) The  number of 
potential ``trials'',
the product of these, is far less than the inverse of the result of 
equation  \ref{fact}.
Therefore, we conclude with very high confidence that
EGRET is detecting radio loud AGN.
We note that this result does not depend on any {\it a priori}
assumption about the emission of \g s by AGN.

\vskip 0.5cm
\centerline{\bf 3.2 The {\bf \it A Priori} Probability of EGRET Detecting a Radio
Source}

Now that we have clearly established that EGRET is detecting some
radio-loud AGN,
it is sensible to estimate
the {\it a priori} probability, $\eta$,
for EGRET to detect a particular flat-spectrum radio source.
This will allow
the   use of  equation \ref{bayes}
to estimate the {\it a posteriori}
 probability
of a correct identification of any specific source.
We do not attempt here to model the effect on $\eta$ of
  the variation across the sky of the depth of the
EGRET exposure. We expect that this cannot effectively be done
because of the large uncertainty of the  
gamma-ray luminosity function of AGN (Chiang \etal\ 1995), but that it is
not a huge effect.
Also,
we assume that $\eta$ is not dependent on the radio spectral index, 
although we  derive $\eta$ for $\alpha>-0.5$. It is appropriate to thus
give sources with $\alpha\le-0.5$ the ``benefit of the doubt'' because
of the potential error in $\alpha$ obtained from the surveys.

We seek a functional dependence on  radio flux for $\eta$ which
is self-consistent as  done by de Ruiter \etal\ (1977) in their
identification of optical counterparts of radio sources.
Specifically,
we solve for a value of $\eta$ such that the integral  of
$p(id|r)$ given by equation \ref{bayes} 
(for radio sources with $\alpha>-0.5$)
divided by the number of radio
sources considered (with  $\alpha>-0.5$)
yields the assumed value of $\eta$.
We analyze $\eta$ for several $S_5$ intervals. These intervals
are chosen to provide adequate statistics while preserving sensitivity to
the $S_5$ dependence of $\eta$.

For the flux density range $S_5>$ 5 Jy, 
iteration leads to the self-consistent value of $\eta=0.19\pm0.08$
with $\Sigma p(id|r)=6.0$ for 32 radio sources.
For the flux density range $2<S_5<5$ Jy, 
iteration leads to the self-consistent value of $\eta=0.13\pm0.04$
with $\Sigma p(id|r)=14.7$ for 109 radio sources.
For the flux density range $1<S_5<2$ Jy, 
iteration leads to the self-consistent value of $\eta=0.034\pm0.011$
with $\Sigma p(id|r)=9.1$ for 266 radio sources.
And, for the flux density range $0.6<S_5<1$ Jy, 
iteration leads to the self-consistent value of $\eta=0.0047\pm0.0032$
with $\Sigma p(id|r)=2.2$ for 473 radio sources. The uncertainty of
$\eta$ is simply assumed to be  $\eta [\Sigma p(id|r)]^{-1/2}$. This is the 
approximate expectation
 for large $S_5$ because of Poisson fluctuation. For small $S_5$, 
it might be an overestimate, but
we expect that it is reasonable given potential systematic error.

We have chosen an analytical representation for $\eta$
which has appropriate limiting values for both high and low values
of $S_5$. The high limiting value is chosen to be consistent with
$\eta=0.19\pm0.08$ for $S_5>$ 5 Jy. An additional two
parameters have been fit to match the relatively well determined
values of $\eta$ 
for $1<S_5<2$ Jy and  $2<S_5<5$ Jy. 
\begin{equation}
\hat\eta(S_5)=0.2 (1-e^{-0.07(S_5/{\rm Jy})^{2.3} } )
\label{eta}\end{equation}
This fit is shown along with the data in Figure 2.
It is slightly in excess of the data for $0.6<S_5<1$ Jy, but
only by a small amount relative to the uncertainty. It is not inappropriate
to thus give small $S_5$ sources the ``benefit of the doubt''
because of variability.
We have also done this analysis for an alternate choice of $S_5$ bins:
$0.5<S_5<0.8$ Jy, $0.8<S_5<1.5$ Jy, $1.5<S_5<4$ Jy, \& $S_5>$ 4 Jy.
The values of  $\eta$ thus obtained are consistent with equation \ref{eta}.
Also, the value of $\eta$ for $S_5$ bins a factor of 2 narrower is consistent
with equation \ref{eta}. Thus, the derived {\it a priori} probability is not
strongly dependent on binning.
In \S 4.5, we discuss the strong probability that the very low
value of $\eta$ for $0.6<S_5<1$ Jy implies a correlation between  
\g\ flux and radio flux.

\begin{figure}
\hbox{\psfig{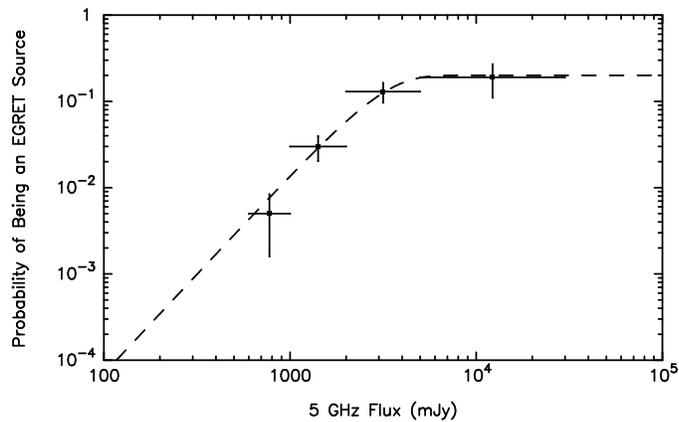}}
\caption{
\baselineskip=11pt 
The measured  {\it a priori} probability $\eta=p(id)$
of a radio source being detected by EGRET as a function of $S_5$.
The fit of equation \ref{eta} is also shown.
} 
\end{figure}

\vskip 0.5cm
\centerline{\bf 3.3 Verification of Previous Identifications}

With the method of \S 2  and the prescription for
an {\it a priori} probability of equation \ref{eta}, we can 
now  look objectively at the radio source identifications of TH95
which are noted in column 4 of Table 1. 
Most of the   ``high-confidence identifications'' of TH95 are indeed so.
Of these 40 identifications;
 six have $p(id|r)\ge0.999$,
seven have $0.99\le p(id|r)<0.999$, nine
have $0.90\le p(id|r)<0.99$, ten have
$0.50\le p(id|r)<0.90$, and six
have $0.10\le p(id|r)<0.50$. 
Only two have $p(id|r)<0.10$:
2356+196 and 1604+159.
The 2EG J0000+2041 EGRET 95\% confidence position
contour contains 3.2 square degrees and radio 
sources as flat and bright as PKS 2356+196 
have a number density of 0.01 per square degree. We find that the fraction
 of flat-spectrum radio sources with the 5 GHz flux density of
PKS 2356+196 (700 mJy) which are detected by EGRET is only about 0.006. Thus we find
with 93\% confidence that PKS 2356+196 is a confusing source rather than the
identification of 2EG J0000+2041.
It is certainly possible that 2EG J0000+2041 is PKS 2356+196, but we suggest
that it should not be considered to be a ``high-confidence identification''.
Similarly, PKS 1604+159 is also a weak 5 GHz sources in a large EGRET
error ellipse.

Four of the ten ``lower-confidence identifications'' of TH95
are  found to be  very unlikely with $p(id|r)<0.005$.
All are modest 5 GHz sources well beyond the 95\% contour of large EGRET
error ellipses. 
On the other hand,
two of the ten ``lower-confidence identifications'' of TH95
are  found to be quite likely.
With $p(id|r)\ge0.95$, the identifications of 0458-020
and 0234+285 are found to be better than 55\% of the ``high-confidence identifications''. 
Both are bright, flat-spectrum
5 GHz sources which have a low number density and the fraction of 
sources this bright at 5 GHz which are detected by EGRET is large.

 Although TH95 classify PKS 1406-076 as a ``high-confidence
identification'', their method would also allow it to be classified
as  a ``lower-confidence'' identification because
PKS 1406-076 lies right on the 95\% confidence contour
of TH95 (see the ApJ CD ROM).
We find that the identification of PKS 1406-076
is correct with 86\% confidence.
Although  PKS 1406-076 is not near the center of the error region, the 
95\% confidence position
contour contains only 0.2 square degrees. 
The variability of the EGRET flux and the
finding of a possible correlation
of the variability of optical flux with the \g\ flux (Wagner \etal\ 1995)
indicates strongly that the identification is correct.  

Table 2 shows the analysis of \S2 for the sources in the
Supplement to the Second EGRET Catalog (Thompson \etal\ 1996). The three
AGN identifications in the Supplement to the Second EGRET Catalog
are found to be compelling. 

\vskip 1cm
\centerline {\bf 3.3.1 The VLBI Flux Density of EGRET Blazars}

We have examined the  2.29 GHz VLBI flux densities (Preston \etal\ 1985)
for the sources in Tables 1\&2  which either have $p(id|r)\ge0.9$,
or have 
$p(id|r)\ge0.7$ for a EGRET source which is highly variable (PKS 1406$-$076)
or for which the confidence of the
identification is otherwise enhanced (Mrk 421). Thirty of these 31
sources (all except PKS 1830$-$210) were measured by Preston \etal\ (1985).
The 30  EGRET
blazars are shown in comparison to the parent population in Figure 3.
The marginal distribution of the VLBI flux density is shown in Figure 4
for the sources in Figure 3 with $\alpha\ge-0.5$.
It is immediately apparent that  the EGRET blazars have 
higher  VLBI flux densities than the parent population. 
A Kolmogorov-Smirnov (KS) test shows that these distributions differ with
99.9994\% confidence. VLBI upper limits were treated as detections 
in this KS test. This is   conservative because  Preston \etal\ (1985) obtained
VLBI detections for all 30 of the EGRET blazars they observed.

\begin{figure}
\hbox{\psfig{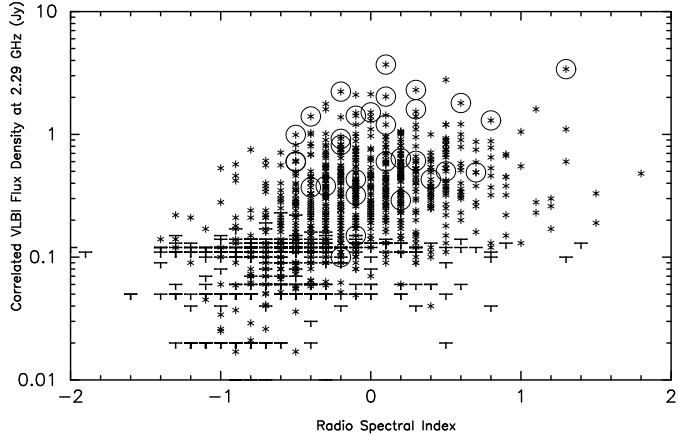}}
\caption{
\baselineskip=11pt 
A scatter plot of the 2.29 GHz VLBI flux densities and radio spectral
indices from Preston \etal\ (1985) for the 1259 of 1398
sources in their catalog with
radio spectral indices. Upper limits for VLBI flux densities are shown for
430 of these. Those which are robustly identified with EGRET sources ($p(id|r)>0.9$, or
$p(id|r)>0.7$ for PKS 1406$-$076 and Mrk 421) are circled. The EGRET blazar shown
with $\alpha=+1.3$ is 3C 454.3 --- the index of Preston \etal\ (1985)
probably differs from the GB survey value $\alpha=+0.1$ because of variability.
} 
\end{figure}
\begin{figure}
\hbox{\psfig{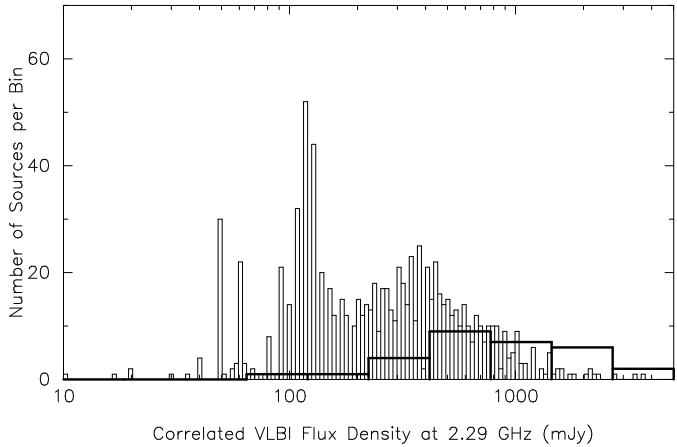}}
\caption{
\baselineskip=11pt 
The marginal distribution of the VLBI flux densities
of Figure 3 for $\alpha\ge-0.5$. The 882 sources in the catalog 
of Preston \etal\ (1985) with $\alpha\ge-0.5$ are shown with the histogram
using thin lines.
Upper limits are treated as detections (at the value of the upper limit). 
The peak at 
\ap120 mJy results from this treatment of upper limits. 
The 30 EGRET blazars circled in
Figure 3 are shown with the histogram using bold lines.
} 
\end{figure}

We note that
the amount of VLBI flux density is a much better discriminant
of EGRET blazars than the VLBI core visibility. 
Figure 5 shows the distribution of the visibilities from
the catalog of Preston \etal\ (1985) for the EGRET blazars. 
A KS test indicates with only 81\% confidence that the
EGRET blazars have higher visibilities than the parent population.
The VLBI visibility is influenced by the amount of flux from the
radio lobes. If the properties of radio lobes do not reflect whether
\g s are visible, it is
expected that the variance of  lobe radio luminosity will cause the observed
effect.

\begin{figure}
\hbox{\psfig{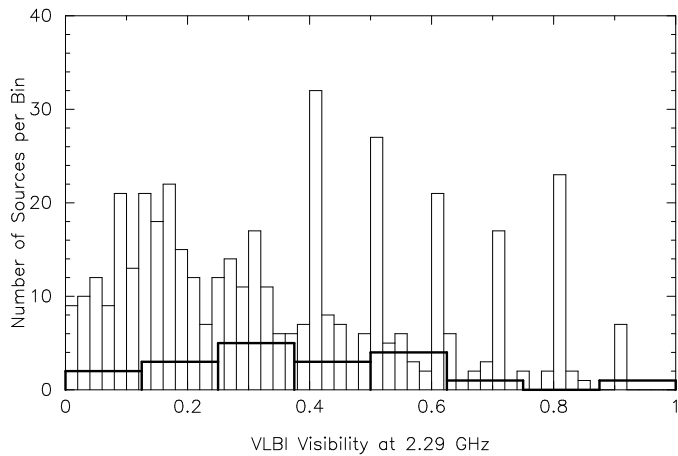}}
\caption{
\baselineskip=11pt 
The distribution of the visibilities (VLBI flux density/total
flux density). The 451 sources in the catalog 
of Preston \etal\ (1985) with $\alpha\ge-0.5$ and a visibility
are shown with the histogram using thin lines.
Upper limits are treated as detections at the upper limit of the visibility.
The 19 EGRET blazars with visibilities in the catalog 
of Preston \etal\ (1985)
are shown with the histogram using bold lines.
} 
\end{figure}

This is the first report of a radio property which distinguishes
the EGRET blazars from the parent population
of flat-spectrum radio sources. It supports
arguments that the \g\ emission
is taking place at the base of a relativistic jet.
This finding also has a practical application.
A large VLBI flux density  enhances the
confidence of an EGRET blazar identification. 

\vskip 1cm
\centerline {\bf 3.4 Identification Analysis of Previously Unidentified EGRET Sources}

An obvious hypothesis is that some of the unidentified EGRET sources
are also blazar type AGN. Our results are shown in
Table 1 for the unidentified EGRET sources in the
Second EGRET Catalog (TH95) and in Table 2 for the
unidentified EGRET sources in the Supplement to the Second EGRET Catalog 
(Thompson \etal\ 1996). We find evidence of additional blazar identifications.
The strongest new identification is
lensed blazar PKS 1830$-$210 near the direction of the Galactic center with
$p(id|r)=0.98$. A total of
four new potential identifications with $p(id|r)>0.5$ appear in Tables
1 \& 2.
They include the \ap2 Jy sources J1650$-$5044 and J1802$-$3940
which are within 10\deg\ of the Galactic plane and 
have not yet been sufficiently studied to be certain that they are not
Galactic sources.
These 4 likely identifications
are given in Table 4 along with an additional 11 possible
identifications with $p(id|r)\ge$0.05. Two additional sources are also
included in Table 4 which have 
0.02$<p(id|r)<$0.05 and an indication of variability with confidence 
in excess of 99\%  (McLaughlin \etal\ 1996) ---
variable EGRET sources are more likely to be blazars.

\vskip 1cm
\centerline{\bf 3.5 Summary of Blazar Identifications}

Table 3 gives the 42 EGRET blazar identifications which we expect to be robust.
All of these sources have been established to be AGN.
Most of these identifications have $p(id|r)\ge0.9$. Several
sources are included with 0.3$<p(id|r)<$0.9 if there is high confidence
of  variability of the EGRET source  (in excess of 99\%), or if a large
VLBI flux density has been observed. Our minimum requirement is 300 mJy in the 
2.29 GHz survey of Preston \etal\ (1985) or 
in a subsequent VLBI observation (details given in the footnotes of Table 3).
${}$From Figure 4, the requirement of a 2.29 GHz VLBI flux density $>$ 300
mJy excludes 54\% of potential confusing sources. 
Table 3 includes 33  of the 40 ``high-confidence identifications''
 of TH95, 2  of 10 ``lower-confidence identifications'' of TH95,
1 new identification of an
unidentified TH95 source, 3 identifications
from Thompson \etal\ 1996 (one of these, NRAO 190, was first
identified by McGlynn  \etal\ 1996), and 3
identifications from subsequent 
EGRET exposures. Table 4 comprises the potential identifications for which
we expect further investigation will be most useful.
Four of these potential identifications are very compelling  with $p(id|r)>0.5$.
Some sources  warrant specific comments.

\begin{itemize}
\item[]
{\bf B0115+0242}:
EGRET  source  2EG  J0119+0312 is identified with 6\% confidence
with radio source  B0115+0242.
Because there is a 92\% confidence indication of variability for 2EG  J0119+0312,
this identification is actually more likely than 6\%.
The optical magnitude is 17.5. The red shift is 0.672. It shows OVV
behavior. This source is also known as PKS 0115+02 and 3C 37.
The  $\alpha_{1.4-4.85}$ radio spectral index is -0.7. 
It would be worthwhile to conduct more
radio and optical flux monitoring observations of this source to 
obtain a radio spectral index from simultaneous observations, to
characterize
variability, and to determine the amount of mm radio flux density. Also, it
would be worthwhile to observe 
the parsec scale radio structure of this source, and
determine a spectral index of possible compact radio emission.

\item[]
{\bf 0219+4248 (3C 66A)}: We find $p(id|r)$ of 4.0\% for the identification
of EGRET source J0220+4228 with  3C 66A assuming a radio spectral index of
$-$0.5. However, a radio index of \ap0. is indicated by 22 and 37 GHz 
monitoring (Ter\"asranta private communication). Using this radio
spectral index,  $p(id|r)$ is found to increase from 4.0\% to 6.2\%. 
We note that 3C 66A is
a BL Lac type blazar which shows optical variability and that the 
 VLBI flux density at 2.29 GHz from Preston \etal\ (1985) is 0.27$\pm$0.05 Jy.
Thus, this identification is credible. However, Verbund \etal\ (1996) 
identify this source with binary ms pulsar J0218+4232 after finding \ap2$\sigma$
evidence of modulation of the EGRET flux after epoch folding.

\item[]
{\bf 0336$-$019}: Source
0336$-$019 (CTA 26) was detected by EGRET  during the interval 05/09/95 --- 05/23/95
(Mattox \etal\ 1995).
The EGRET 95\% confidence contour is well fit by an ellipse centered on
l=188.25, b= $-$42.34, and with  semi-major axis $a$=44\min, 
semi-minor axis $b$= 37\min, and $\phi$=  27\deg.
A complete description of multiwavelength observation is in preparation.

\item[]
{\bf 0446+112}: This identification by TH95 is plausible, but narrowly fails to meet our
criteria with $p(id|r)=0.31$ and S$_{\rm VLBI}=200\pm20$ mJy. 
It is possible that this 2.29 GHz VLBI flux density measurement (Preston \etal\ 1985)
is under representative because of variability.

\item[]
{\bf 1229$-$021}: This identification by TH95 is plausible, but fails to meet our
criteria with $p(id|r)=0.32$ and S$_{\rm VLBI}=90\pm30$ mJy.
It is possible that this 2.29 GHz VLBI flux density measurement (Preston \etal\ 1985)
is under representative because of variability.

\item[]
{\bf B1324+2226}: 
EGRET  source  2EG  J1324+2210 is identified with 63\% confidence
with this GB source.
The  $\alpha_{1.4-4.85}$ radio spectral index indicated by the
(non-simultaneous) GB survey flux densities is +1, very strongly 
inverted. Recent simultaneous cm radio measurements indicate a radio
spectral index of \ap0. (M. Aller, private communication).
A VLBI
position was published by  Johnston \etal\ (1995) but a VLBI flux was not
given.

\item[]
{\bf J1650$-$5044}: Our analysis shows a
strong probability (92\%) that this PMN source is
EGRET source 2EG J1648$-$5042. However, 
observations at other wavelengths are
required to ascertain that this radio source is an AGN rather than a Galactic
radio source. This source was previously detected as radio
source MRC 1646$-$506. Otherwise, it does not appear in the literature.
An observation with the Australian Telescope Compact Array would be useful
 (it is too far south for the VLA). 

\item[]
{\bf J1802$-$3940}: 2EG J1800-4005  is identified with 
PMN  J1802$-$3940 with 81\% confidence.
Observations at other wavelengths are
required to ascertain that this radio source is an AGN rather than a Galactic
radio source. 

\item[]
{\bf 1830$-$210}: 2EG J1834$-$2138
 is identified with PKS 1830$-$210 with 98\% confidence. No VLBI
observation has been reported for this source.
This 8 Jy source has not  been identified optically (it is
6\deg\ from the Galactic disk near the Galactic
center). A VLA radio image shows what appears to be
a steep-spectrum 
Einstein Ring and two parity-reversed images of a flat-spectrum core-jet structure
separated by 1\min\min\ (Jauncey \etal\ 1991).
This structure is interpreted as a gravitationally
lensed AGN jet. Thus, it could be an EGRET blazar.
Wiklind and Combes (1996) established through mm observations of hydrocarbon absorption lines 
that the redshift of the lensing galaxy is  z=0.89. If a detailed lens
model can be developed, the autocorrelation function of future 
high-temporal density  
\g\ flux measurements may reveal information about the location of the 
\g\ emitting region. The expected time delay between the two components
is \ap\pt7 seconds (Jauncey \etal\ 1991).

\item[]
{\bf 1908$-$201}:
PKS 1908$-$201 is identified with 2EG J1911$-$1945 with 92\% confidence.
This 2 Jy source has not yet been identified optically (it is
21\deg\ from the Galactic center).
The 8.4 GHz flux density of PKS 1908$-$201 is 2.3 Jy
(Wright \etal\ 1991). Using the consequent index $\alpha=0$
(instead of assuming $\alpha=-0.5$ without a 2.7 GHz detection)
increases the probability of the identification to \ap95\%. 
The correlated VLBI 2.29 GHz flux density is 0.46$\pm$0.05 Jy
(Preston \etal\ 1985).
It would be worthwhile to observe the 
the parsec scale radio structure of this source.
An   optical (or IR) identification and spectrum is also of interest.
Preston \etal\ (1985) give a position which is accurate to 1\min\min:
B1950, 19 8 12.6 $-$20 11 57. or J2000, 19 11 09.8 $-$20 06 57.

\item[]
{\bf 1933$-$400}: This identification by TH95 is plausible, but narrowly fails to meet our
criteria with $p(id|r)=0.34$ and S$_{\rm VLBI}=280\pm30$ mJy.
It is possible that this 2.29 GHz VLBI flux density measurement (Preston \etal\ 1985)
is under representative because of variability.

\item[]
{\bf 2022$-$077}: This identification by TH95 is plausible, but fails to meet our
criteria with $p(id|r)=0.61$ and S$_{\rm VLBI}=130\pm20$ mJy.
It is possible that this 2.29 GHz VLBI flux density measurement (Preston \etal\ 1985)
is under representative because of variability.

\item[]
{\bf 2155$-$304}: An EGRET detection of this X-ray selected BL Lac type
object was reported by Vestrand, Stacy and Sreekumar (1995). Since these
authors do not provide a specification of the 95\% confidence source location
region, a likelihood analysis (Mattox \etal\ 1996a) of
 the VP 404 data for E$>$ 100 MeV was done. The 
 95\% confidence contour is well fit by an ellipse centered on
l=17.26 , b= $-$52.28, and with  semi-major axis $a$=45\min, 
semi-minor axis $b$= 40\min, and $\phi$=  79\deg.
The analysis of \S 2 indicates that B2155$-$304 ($S_5$=407 mJy, $\alpha$ = 0.3)
can be identified with the EGRET source with $p(id|r)$=12\% confidence. A 
test for variability of the EGRET source using the average flux of
phase 1\&2 of the mission, VP 209, and VP 404 results in $\chi^2$=
7.3 for 2 DOF. Thus, a
constant \g\ flux  can only be rejected at the 97.5\% confidence level. 
Preston \etal\ (1985) measured a 2.29 GHz VLBI flux density of 144$\pm$8 mJy.
The identification is very plausible, but fails to meet our
criteria  for inclusion in Table 3.

\item[]
{\bf 2200+420}: A 4.4$\sigma$ EGRET detection of 2200+420 (BL Lacertae)
is reported by Catanese \etal\ (1997). 
The 
 95\% confidence contour for the event selection E$>$ 100 MeV
is well fit by an ellipse centered on
l=92.59 , b= $-$10.44, and with  semi-major axis $a$=89\min, 
semi-minor axis $b$= 58\min, and $\phi$=  35\deg.
The analysis of \S 2 indicates that B2200+420 ($S_5$=3.6 Jy, $\alpha = -0.2$)
can be identified with the EGRET source with $p(id|r)$=99.3\% confidence.
A high confidence identification is obtained in spite of the large error 
region because the 5 GHz flux density is large.
\end{itemize}

\vskip 1cm
\centerline{\bf 3.6 Limitations of this Analysis}

Although it is a significant improvement over previous EGRET identification
work,
we note that our technique is subject to a number of 
potential sources of systematic error.
The  error ellipses of TH95 indicate  statistical error only. Substantial
systematic error is also expected, but was not given because of the
difficulty of quantifying it.
As discussed by TH95 and Mattox \etal\ (1996a),
error in the model of the Galactic diffuse \g\ emission and undetected
nearby \g\ point sources can have a substantial effect on  EGRET
position estimates.
Because the present investigation is limited to  $|b|>3$\deg,
the effect of error in
the Galactic diffuse model of Hunter \etal\ (1996)
is not severe for the EGRET sources we study in this paper.

The radio flux densities of blazars vary by  
factors as large as \ap10. Therefore,
it is problematic to use the radio survey flux 
for the identification analysis because the flux at the survey epoch
could differ significantly from the average flux. Also, the radio
spectral index we use  was obtained from non-simultaneous surveys
at 4.85 GHz and 1.4 GHz. It is subject to error because of flux variability. 
The error that this introduces can be mitigated by follow-up radio
observations of sources
which are potential EGRET source identifications.
For this reason, potential identifications with
values of $p(id|r)$ as small as \ap0.05 should not be dismissed until 
follow-up observations have been made.

The necessary simplification of assuming that
the {\it a priori} probability,
$\eta$,  is independent of  
EGRET exposure and  radio spectral index
also contributes some error to our
estimates of the probabilities of correct identifications, but
we expect that  they are not severely compromise.
An additional minor error in our identification analysis may  result from
some deviation of the dependence  of the
log of the likelihood of the EGRET
data on the source position from the assumed paraboloid.
Also, a minor error in our identification analysis is introduced
by the discernible but small differences between the elliptical fits and the
actual  position confidence contours of TH95 which are
available on the  ApJ CD ROM corresponding to TH95.

For these reasons, the probabilities of correct identification derived
here are indicative rather than definitive. We expect that for identifications
with an {\it a posteriori} probability $p(id|r)>0.9$, 1$-p(id|r)$  could
be in error by as much as a factor of 5. Thus, 
$p(id|r)=0.9$ implies $0.5<p(id|r)<0.98$. Similarly we expect 
that for identifications
with an {\it a posteriori} probability $p(id|r)<0.1$, $p(id|r)$  could also
be in error by as much as a factor of 5. Thus,
$p(id|r)=0.1$ implies $0.02<p(id|r)<0.5$. Although the large uncertainty
of $p(id|r)$ is not ideal, it is not so large as to render this analysis
useless. We expect that our method offers an order of magnitude improvement
in precision over that of TH95.

For latitudes less than \ap15\deg, a substantial
fraction of the radio sources will  be Galactic radio sources. 
We search
for counterparts in this range  because correct blazar identifications
could result nonetheless. 
In fact,
one planetary nebulae and  one H$_{II}$ region 
appear in Table 1 (and are noted in the footnotes), but are not thought
to be \g\ source counterparts.
We find that no excess
of compelling identifications results at low  latitudes. Therefore, undiscerned
Galactic  radio sources are apparently not a major problem for Galactic
latitudes   3\deg$<|b|<$15\deg.

We do not quantitatively incorporate  in our identification analysis
 the variability of EGRET sources, nor the amount of
 correlated VLBI flux density --- although these
 certainly pertain. They are considered in the selection of the
high probability EGRET identifications of Table 3 and the ``most
interesting'' potential identifications of Table 4.

Also, there is potentially more information for some sources which would
be difficult to quantitatively incorporate  in our identification analysis.
We note that the analysis described in \S 2
results in a formal probability of 70\% that MRK 421 is correctly identified
as an EGRET source. This results from  the low radio flux of MRK 421, 
and the large EGRET position uncertainty for this weak source. Additional information
drastically improves the confidence of this identification.
 MRK 421  has 
distinguishing characteristics which effectively reduce the number 
density of such sources.
In addition to a substantial correlated VLBI flux density and
 domination of the optical
emission by  continuum synchrotron radiation,
MRK 421 shows a  large and highly variable X-ray
flux. It 
has been detected at TeV energies through the observation of \v Cerenkov
light produced in the atmosphere by \g\ induced electromagnetic cascades
(Buckley \etal\ 1996).
The TeV flux is highly
variable and it has been observed to vary in correlation with the
X-ray flux (Buckley \etal\ 1996).
The TeV identification is very secure because the position of
the TeV emission has been determined to \ap4\min\ and is consistent 
with the MRK 421 position (Buckley \etal\ 1996).

The identifications of Table 1 indicate that the
OVV property is common to many of the sources detected
by EGRET which have been frequently observed with optical telescopes.
Thus, the OVV property may serve to distinguish sources and improve
identification confidence. This may also be true for optical polarization and
BL Lac classification.
However, careful work remains to be done to ascertain
 that these characteristics distinguish EGRET
sources. 
Wagner (in preparation) finds that 80\%  of
all sources which have a flat radio spectrum up to 90 GHz are OVV
irregardless of whether or not they are detected by EGRET. 
The fraction of EGRET sources which are believed to be
OVV is not significantly
larger than this; however not all have been yet been extensively monitored
in the optical band.
Also, the fraction of blazars detected by EGRET which are radio-selected
BL Lac type objects
is not significantly different from the fraction of flat-spectrum
radio sources which are BL Lacs.
Also, optical polarization
may be generic for flat-spectrum radio sources.
Fugmann (1988) has studied
optical polarization observations of the K\"uhr \etal\ (1981) sample.
He finds
that the probability of measuring
significant optical polarization for a
flat-spectrum radio source in a single measurement is
62\%. It is possible that all flat-spectrum
radio sources would have the property of high optical polarization in at least
one observation if sufficient observations were made.

\vskip 1cm
\centerline{\bf 4. Discussion}

We find that a substantial number
 of the identifications of TH95 are far from
certain. This  has implications for most studies of the
``EGRET blazars''. It impacts the study of the \g\ luminosity function
and evolution (Chiang \etal\ 1995). It should be considered in studying
the correlation of \g\ and radio luminosity (Padovani \etal\ 1993; 
Stecker, Salamon, and Malkan 1993;  Dondi and Ghisellini 1995;
M\"ucke \etal\ 1996), and in the use of the EGRET detections in
calculations of the production of 
isotropic diffuse high-energy \g\ background  by
AGN (Padovani \etal\ 1993; Stecker \etal\ 1993; 
Salamon and Stecker 1994;  Chiang \etal\ 1995;
Stecker and Salamon 1996; Kazanas and Perlman 1996). Careful 
consideration of the confidence of the identification is also in order
when seeking  distinguishing properties of the ``EGRET blazars''. Unless
explicit consideration is given to potential misidentification, we suggest
that these analyses be restricted to the 42 EGRET blazars of Table 3
for which $p(id|r)$ exceeds 90\%, or a high confidence identification is
otherwise indicated. We estimate that \ap1 of these identifications is 
specious.
Additional EGRET exposure and additional
analysis and observation at other wavelengths
may eventually contribute \ap10 more sources to this population.

\vskip .5cm
\centerline{\bf 4.1 On the Detection  of Blazars by EGRET}

We have demonstrated in \S 3.1 with confidence far beyond question
that EGRET is detecting radio loud AGN.
Six of the identifications in Table 1
have a probability of being correct of at least 99.9\% (0420$-$014, 0528+134,
3C 273, 3C 279, 1633+382, and 3C 454.3)
All of these
have flat radio spectra ($\alpha\ge-0.5$). All have substantial correlated VLBI flux density.
All 6 show significant changes in optical flux on short time-scales.
Of these 6, the  4 which have been extensively observed
show significant optical polarization.
Also, the 5 which have been observed sufficiently with VLBI to discern it show
superluminal motion of the parsec scale structure of radio jets.
Therefore, it is apparent that EGRET is detecting the blazar class of AGN.

The other identifications of Table 3 all have radio spectra with
$\alpha\ge-0.5$, except for 0521$-$365 for which
$\alpha= -0.7$ from the PMN catalog. 
This source has not been extensively observed. The report of
an EGRET detection lead to VLBI observations which 
reveal that the spectrum of the core is $\alpha_{4.9-8.4}=+0.3$
 (Tingay \etal\ 1996). Thus, if this source is truly steep-spectrum
in single dish observations,
it is due to the flux of the radio lobes.
Therefore, it appears
that all  high confidence EGRET identifications  
are  flat-spectrum radio sources. 
Fugmann (1988) concludes from his study of the optical polarization
that at least 2/3 of all flat-spectrum radio sources
are HPQ sources and therefore blazars. 
Therefore, it is plausible that all extragalactic  EGRET source
 are blazars.

It is interesting that
EGRET is detecting only \ap10\% of the blazars. It is possible
that this results from time variability of the gamma-ray emission. However,
a ``duty cycle'' of \ap30\%  is observed for the strong EGRET sources --- too
large to explain the fact that only ~10\% of the blazars are detected,
unless there is strong variability on time scales longer than \ap1 year.
Another possibility is that the gamma-ray luminous blazars differ intrinsically
from the blazar parent population. E.g., the gamma-ray emission may be more
narrowly beamed than the longer wavelength emissions so that the \g\
blazar sub-population has a smaller angle between the jet and the line of
sight than the blazar parent-population (Salamon \& Stecker 1994, Dermer 1995).

\vskip .5cm
\centerline{\bf 4.2 Non-Detection of Steep Radio Spectrum and Radio Quiet AGN
by EGRET }

In Figure 6 we present a scatter plot of $S_5$ and $\alpha$ for
all radio sources considered. The identifications of Table 1 \& 2
with $p(id|r)\ge 0.02$ are surrounded by circles. The size of the circle
increases with the confidence of the identification.
It appears that the distribution of observed spectral indices for the identified sources
is consistent with that of  flat-spectrum  sources ($\alpha>-0.5$)
given the errors in the spectral indices due to the non-simultaneity 
of the radio surveys. Thus, it is plausible that EGRET is
detecting only flat-spectrum radio sources.

\begin{figure}
\hbox{\psfig{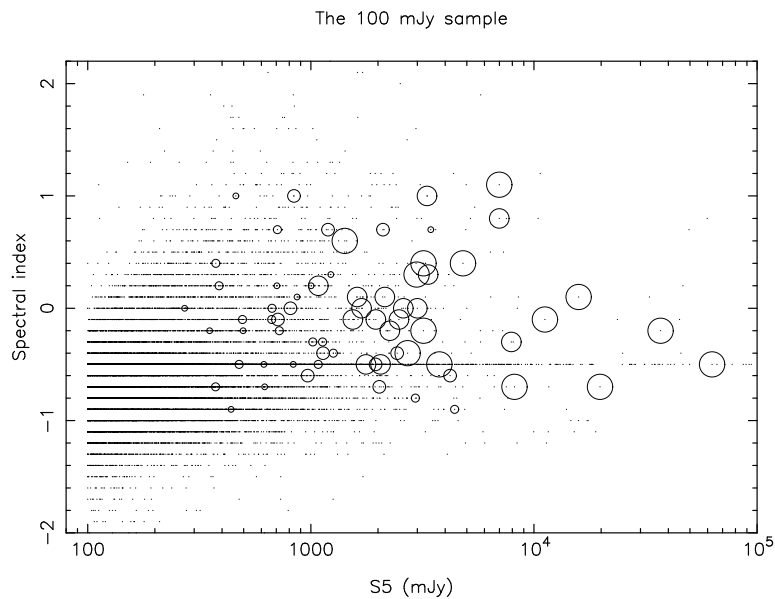}}
\caption{
\baselineskip=11pt 
The distribution of radio flux density ($S_5$) 
and  radio spectral index ($\alpha$) for all
sources in  the PMN and GB surveys with $S_5> 100$ mJy. We plot  an 
index of $\alpha=-0.5$ if an index is not available from the survey catalogs
(as assumed for the analysis of \S 2).
The consequent enhanced source
 density at $\alpha=-0.5$ is apparent.
The identifications of Table 1 and Table 2
with $p(id|r)\ge 0.02$ are surrounded by circles (using $\alpha$  from Table 1
or 2).
The size of the circle
increases  with $p(id|r)$. The five different sizes correspond to the following
ranges: 
0.02 $\le p(id|r) <$ 0.1,
0.1 $\le p(id|r) <$ 0.5,
0.5 $\le p(id|r) <$ 0.9,
0.9 $\le p(id|r) <$ 0.99,
0.99 $\le p(id|r) <$ 1.00.
} 
\end{figure}

It is a possible  concern that the non-detection of steep-spectrum sources
results from our use of the reduced number density of
flat-spectrum radio sources in the calculation of the likelihood ratio.
Therefore, we have repeated the analysis described in \S 2
with a  number density of radio sources
$\rho(S_5)$ which depends only on flux density. This results in values
of $LR$ smaller by a factor of \ap2. However, significant detections
persist and the  list of strong identifications is nearly unchanged.
 The sum for the sources in Table 1 is then
$\Sigma p(id|r)=31.7$ (down from 36.6 with the
radio index dependent number density described in section 2). 
In the null hypothesis,
the Monte Carlo calculation
with random EGRET positions  described in section 3.1 yields
$\Sigma p(id|r)=4\pm2$. 
Thus, the result that
EGRET is detecting  radio loud AGN does not depend
upon the assumption that the flat spectrum sources are more likely
to be detected. It follows that the steep-spectrum sources
are generally not being  detected. Therefore,  the
use of the spectral index in the  calculation of
confusing source number density is justified.

We also suggest that
EGRET is not detecting radio-quiet AGN. 
If an unidentified EGRET source
is actually a blazar, that blazar is very
likely to be among the possible identifications
given here.
If the EGRET flux density is variable,
the likelihood of a blazar  identification is enhanced.
We note that every variable EGRET source in Table 1
(with variability index $\ge2$ corresponding to
confidence of variability $\ge$99\%), has a plausible radio-loud
identification. If we assume that the \g\ emission of all AGNs would vary,
we can conclude that
EGRET is not detecting any  radio-quiet AGN. 
We note that there are several unidentified EGRET sources at $|b|<3$\deg\
which are excluded from our analysis which are probably Galactic (e.g.
Tavani \etal\ 1997).

\vskip .5cm
\centerline{\bf 4.3 Suggested Observations at Other Wavelengths of Potential
Gamma-Ray Selected Blazars}

Although we have found only one new EGRET identification in this analysis 
(PKS 1830$-$210)
which qualifies for inclusion in the
high confidence  identifications of Table 3,
it is apparent that a substantial number of the previously
unidentified EGRET sources are also blazars. 
Further  study at other wavelengths (primarily
radio and optical) is  required to ascertain
whether or not some of these candidates have  blazar properties.
Most of these potential EGRET sources 
have not been studied extensively.
Upon further investigation, they may be
found to be ``gamma-ray selected blazars''.
In Table 4 we list the  17 ``most interesting'' 
new potential identifications. 
This table also gives the best positions available for these sources.

Further radio observation can establish
variability and a radio spectrum from simultaneous observations.
Radio interferometric images (VLBI, or VLA/AT) can be obtained to establish
that the radio source is an AGN. A large correlated VLBI flux
density is indicative of EGRET sources.
Measurements of flux densities in the mm radio band may help to
distinguish blazars.
Measurements of optical variability and polarization
can be used to establish membership in the OVV or  HPQ blazar class.
Spectra to establish 
redshifts, and equivalent widths of emission lines  are of interest, as
 are X-ray and infrared
  observations.
Simultaneous multiwavelength observation will be of interest
once identifications are firmly established.
In principle, the identification could be secured by observation
of gamma-ray flux variation which correlates with variation at a
longer wavelength. However, the  EGRET sensitivity is generally
not sufficient for this unless the \g\ flux is exceptionally large.

In many cases, an optical counterpart is obvious from an examination
of the POSS (or ESO) plates for the GB or PMN error region. In other cases,
an optical counterpart is not clearly apparent in this \ap20\min\min\ region
because of source
confusion. 
Radio observations to obtain an interferometric position with the
VLA or  the Australian Telescope Compact Array (AT) are then appropriate.
Positions accurate to \ap0.01\min\min\ are given in Table 4
for  sources of likely interest which
were observed by Patnaik \etal\
(1992 \& 1996) with the VLA for
potential use as MERLIN phase calibrators.

\vskip .5cm
\centerline{\bf 4.4 Pulsar Candidates}

The non-variable EGRET
sources for which a radio-loud
identification is not apparent are more likely than other EGRET sources
to be pulsars. Tables 1\&2 may thus be used to 
to select optimal pulsar candidates.
Romani and Yadigaroglu (1995) predict that 70\% of gamma-ray luminous pulsars
are in fact radio-quiet. Direct detection of periodicity 
for more than \ap2 through a FFT of the EGRET data is unlikely (Mattox
\etal\ 1996b).
More work is required to determine if the predicted
spatial distribution of pulsars (Yadigaroglu  and Romani 1996) matches that of
the pulsar candidates of Tables 1\&2.

We note that both EGRET
sources in Table 1 which coincide with SNRs (2EG J0008+7307 and 2EG J0618+2234)
have no flat-spectrum radio identifications with $p(id|r)>2$\tpt{-6}. 
This is consistent with
the claim (Sturner and Dermer 1994, Esposito \etal\ 1996,
Sturner, Mattox, and Dermer 1996) of a correlation between
EGRET sources and SNRs.
This correlation could be either the result of enhanced interaction
of cosmic rays with the interstellar medium near SNR, or  
pulsars associated with the SNRs or with young stellar associations
co-located with a SNR.

\vskip .5cm
\centerline{\bf 4.5 The Correlation of Gamma-ray Flux with Radio Flux }

With the limited EGRET data, the dependence of 
the {\it a priori} probability,
 $\eta$, on $S_5$ cannot be understood in detail. 
A simple hypothesis is that  \ap20\% of all
flat spectrum radio sources emit \g s, and that the observed decrease in
$\eta$ with radio flux  
is the result of detecting a fraction of \g\ blazars which decreases with
radio flux because  \g\ flux is correlated with radio flux and  EGRET 
has a relatively high detection threshold. 
 The initial reports of a correlation between 
\g\ luminosity and radio luminosity (Padovani \etal\ 1993; 
Stecker \etal\ 1993; Salamon and Stecker 1994; Stecker and Salamon 1996)
are flawed in that they fail to consider the effect of a common redshift
on the apparent correlation. An analysis which overcomes this defect
through
partial correlation analysis (Dondi and Ghisellini 1995) indicated
a correlation with 99.5\% confidence. However, a partial correlation analysis 
which also considers upper limits (M\"ucke \etal\ 1996) does not
find significant correlation.
We present
a simple argument below which establishes with 99.998\% confidence that
the maximum observed \g\ flux of EGRET blazars correlates with the
average 5 GHz radio flux.

We simply note that all of the  EGRET blazars which are \g\ bright
are also  bright radio sources. 
The ten EGRET blazars with a peak \g\
flux above \pt{-6}\cms at E$>$100 MeV
(0208-512, CTA 26, 0528+134, 1156+295, 3C 279, 1406$-$076,
1622$-$297, 1633+382, 1730$-$130, and 3C 454.3)
all have an average radio flux density $S_5$ which exceeds 1.0 Jy. 
${}$From the GB and PMN
catalogs, we estimate that 58\%
of the $S_5>$1.0 Jy radio sources have flat spectra ($\alpha\ge-0.5$),
and that there are 183 of them away from the Galactic plane ($|b|>$3\deg). 
The fraction detected as bright \g\ blazars is 0.05$\pm$0.02.
Using Poisson
statistics, a 99.998\% confidence lower limit on the fraction of these
which are bright \g\ sources is 0.0098.
Examining the GB and PMN catalogs over narrow intervals (250 $<S_5<$ 350 mJy,
350 $<S_5<$ 500 mJy, and 500 $<S_5<$ 1000 mJy) and correcting for the
variable fraction of sources
which have measured spectral indices, and the variable 
fraction which are flat spectrum; we estimate that there are 1130
flat spectrum radio sources with 250 mJy $<S_5<$ 1.0 Jy 
away from the Galactic plane ($|b|>$3\deg). 
If the \g\ flux of EGRET sources were not correlated with radio flux, we
would expect 5$\pm$2\% or 62$\pm$20 of these to be bright \g\ sources.
Using the 
99.998\% confidence lower limit on the fraction detected,
at least 11.1 of these radio sources should be 
bright \g\ sources. None are  observed. The Poisson probability
of observing 0 when 11.1 are expected is 1.5\tpt{-5}. 
We thus reject the hypothesis
that the \g\ flux of EGRET sources is not correlated with radio flux
with 99.998\% confidence. It is possible that the correlation
is non-linear as suggested by M\"ucke \etal\ (1996).

This analysis assumes that the distribution of
$\alpha$ is the same for the sources where it was not measured as it is for
 sources where it was measured. This may not be true in detail, but it is
expected to be a good approximation for sources with $S_5>250$ mJy.
We have effectively analyzed the correlation between
the maximum observed \g\ flux of EGRET blazars and the 
average 5 GHz radio flux because that is essentially what the current data
offers. Most  EGRET blazars are only detected by EGRET during a flaring
state. And the GB and PMN surveys with a 5 GHz observation at one epoch
offer approximately an average flux.

If  bright EGRET blazars with 250 mJy $<S_5<$ 1.0 Jy existed, 
the method of \S 2 could  identify them
because the \g\ error regions 
would still be small. We analyzed the 10 bright EGRET sources listed above
with the counterpart radio flux
artificially altered
to $S_5$=250 mJy. All identifications remain above the threshold
for inclusion in Table 1 ($p(id|r)>$0.001), and 9 of 10 above the threshold
for variable EGRET sources in Table 4 ($p(id|r)>$0.02). Two sources
(3C 279 and 0528+134) would still be above the threshold
for variable EGRET sources  in Table 3 ($p(id|r)>$0.3).
This a mute point however. There are no bright EGRET sources
(peak \g\
flux above \pt{-6}\cms, E$>$100 MeV) at high Galactic latitude ($|b|>$3\deg)
which do not coincide with either a pulsar or a radio source with $S_5>$1.0 Jy.
In fact, this argument can be extended to the myriad radio sources with
$S_5<$ 250 mJy. The prevalent interpretation of the isotropic
high-energy \g\ diffuse radiation is that it 
is due to the \g\ emission of these sources (Padovani \etal\ 1993; Stecker \etal\ 1993; 
Salamon and Stecker 1994;  Chiang \etal\ 1995;
Stecker and Salamon 1996; Kazanas and Perlman 1996). If this interpretation
is correct, there must be a correlation between \g\ and radio flux or
else the weakest
radio sources would also be detected as bright \g\ sources.

We note that 
a Spearman rank-order correlation test of the sources in Table 3 for the
 5 GHz radio flux of Table 3 and the peak \g\ flux 
(without doing a K-correction) suggests
a correlation (with 67\% confidence). The same test shows a
correlation with 99.2\% confidence for the peak \g\ flux and correlated
VLBI flux of Table 3. This is consistent with the finding of
\S 3.3.1  that
the  EGRET blazars are distinguished by 
a large VLBI flux density; and
adds more weight to arguments that the \g\ emission
is taking place at the base of a relativistic jet.
A tighter correlation might be obtained using simultaneous \g\ and
VLBI fluxes (perhaps allowing for a delay in VLBI flux because of
synchrotron self-absorption of radio flux at the base of the jet). 
We note that the same analysis with
the non-simultaneous
8.4 GHz VLBI flux densities of the 28
sources in Table 3 which were measured by
Morabito \etal\ (1986)  shows much less correlation (the
Spearman rank-order correlation test indicates a
34\% probability of a correlation). 
To some extent this is caused by the reduced number of sources included
in the correlation analysis. Perhaps it also reflects
higher radio variability at higher frequency.

\vskip .5cm
\centerline{\bf 5. Summary }

We present a means to quantitatively assess the probability that
an identification of an EGRET source with a radio source is correct. We
also demonstrate conclusively that EGRET is detecting \g\ emission from
the blazar class of AGN. Our analysis suggests that EGRET is not detecting
radio quiet and
steep radio spectrum AGN.
This identification method is used to ascertain with 99.998\% confidence
that the \g\ and radio flux of EGRET blazars are correlated.
Previous EGRET identifications are re-examined.
Table 3 gives  the 42 EGRET blazars  for which the confidence of 
a correct identification is high. 
It is found that  the peak \g\ flux and the 
VLBI radio flux of these sources is correlated with 99\% confidence.
One high confidence  AGN identification (PKS 1830$-$210) is found among
the unidentified EGRET sources of TH95. Sixteen additional
potential identifications
are presented in Table 4
with the best available position for study at other wavelengths.

\vskip 1cm
\centerline{Acknowledgments}

This research has made extensive use of
the NASA/IPAC Extragalactic Database
(NED) which is operated by the Jet Propulsion Laboratory, California
Institute of Technology, under contract with the National Aeronautics
and Space Administration.
J. Mattox  acknowledges support from NASA Grants NAG5-2833 and NAG5-3384.

\vfill\eject

\end{document}